\newcommand{\p}[1]{(\ref{#1})}
\documentstyle[12pt]{article}
\def\a{\alpha}

\def\m{\mu}

\def\t{\theta}

\def\T{\Theta}

\def\vf{\varphi}

\def\be{\begin{equation}}
\def\ee{\end{equation}}
\def\arr{\begin{array}{rll}}
\def\ea{\end{array}}
\def\bea{\begin{eqnarray}}
\def\eea{\end{eqnarray}}

\begin{document}
\renewcommand{\thefootnote}{\fnsymbol{footnote}}
\begin{titlepage}
\noindent

\vskip 3.0cm

\begin{center}

{\Large\bf $AdS_2/CFT_1$, Canonical Transformations}\\

\bigskip

{\Large\bf and Superconformal Mechanics}\\

\bigskip

\vskip 1cm

{\large S. Bellucci${}^{a,}$}\footnote{bellucci@lnf.infn.it},
{\large A. Galajinsky${}^{a,b,}$}\footnote{galajin@mph.phtd.tpu.edu.ru},
{\large E. Ivanov${}^{c,}$}\footnote{eivanov@thsun1.jinr.ru} and {\large S. Krivonos${}^{c,}$}\footnote{krivonos@thsun1.jinr.ru}

\vskip 1.0cm

{\it ${}^a$INFN--Laboratori Nazionali di Frascati, C.P. 13,
00044 Frascati, Italy}\\

\vskip 0.2cm
{\it ${}^b$Laboratory of Mathematical Physics, Tomsk Polytechnic University, \\
634050 Tomsk, Lenin Ave. 30, Russian Federation}

\vskip 0.2cm

{\it ${}^c$ Bogoliubov Laboratory of Theoretical Physics, JINR, \\
141980 Dubna, Moscow Region, Russian Federation}

\end{center}
\vskip 1cm

\begin{abstract}
\noindent We propose a simple conformal mechanics model which is classically equivalent to a
charged massive particle propagating near the $AdS_2\times S^2$ horizon of an
extreme Reissner-Nordstr\"om black hole. The equivalence holds for any finite value
of the black hole mass and with both the radial and angular degrees of freedom of the
particle taken into account. It is ensured by the existence of a canonical transformation
in the Hamiltonian formalism. Using this transformation, we construct the Hamiltonian of a
$N=4$ superparticle on $AdS_2\times S^2$ background.
\end{abstract}

\vspace{0.5cm}


\end{titlepage}
\renewcommand{\thefootnote}{\arabic{footnote}}
\setcounter{footnote}0
\noindent
{\bf 1. Introduction}\\

\noindent In the web of $AdS/CFT$ dualities the $AdS_2/CFT_1$ case has a distinguished status
and still remains to be fully understood \cite{rev}.
One of its peculiarities is that in $d=1$ one encounters superconformal algebras which
cannot be obtained by a dimensional reduction from higher dimensions
(see e.g.~\cite{strominger} for a review).
Using this type of the $AdS/CFT$ correspondence one can hope to get insights
into quantum properties of supergravity black holes
studying simple (super)conformal mechanics as the relevant boundary
theory~\cite{kallosh}--\cite{strominger1}.

An interesting application of the $AdS_2/CFT_1$ correspondence is provided
by a massive charged particle propagating near the horizon of an extreme Reissner-Nordstr\"om
black hole~\cite{kallosh}. The geometry characterizing this case is $AdS_2\times S^2$
and in the limit of large black hole mass $M$ one recovers\footnote{To be more precise, one considers a specific
limit when the black hole mass $M$ is large and the difference between the
particle mass and the absolute value of its charge $(\mu-q)$ tends to zero,
with $M^2(\mu-q)$ being kept fixed.} the conformal mechanics of~\cite{aff}.
This relationship~\cite{kallosh} suggested
an elegant resolution for the problem of an infinite number of quantum states of a particle probe
localized near the horizon of a black hole (see the relevant discussion in Ref.~\cite{strominger}).
It was traced to the absence of a ground state in the conformal mechanics and the necessity
of redefining the Hamiltonian~\cite{aff}.

It is important to notice, however, that it is the radial coordinate of $AdS_2\times S^2$
which is identified with the degree of freedom described by the conformal mechanics.
The angular variables effectively decouple in the large $M$ limit and show up only
in an indirect way via the effective coupling constant. The latter point recently received
attention~\cite{ikn}, where a particular case of the general transformation constructed in
\cite{bik} was considered. It was shown that
the radial part of the particle on $AdS_2\times S^2$ background is classically equivalent
to the conformal mechanics for any {\it finite} value of the black hole mass, i.e.
without taking any specific limit.

In order to get further insights into quantum properties of a test particle near
the horizon of a black hole, a proper accounting of the angular degrees of freedom
is necessary. It is the purpose of this letter to construct a simple conformal mechanics,
which is classically equivalent to a particle moving on $AdS_2\times S^2$ background,
with both radial and angular variables being retained. Specifically, we take the advantage
of the Hamiltonian formalism and demonstrate that the two theories are connected
by a {\it canonical} transformation. The clue to finding such a transformation is
offered by the symmetry group. Requiring the conserved charges to coincide in
both theories, one reveals the desired canonical transformation.

The outline of the paper is as follows. In the next section we compare the radial part
of the particle on $AdS_2\times S^2$ with the conformal mechanics of Ref.~\cite{aff}.
Equating the conformal currents (which involve the Hamiltonian!) inherent both theories
we find a canonical transformation which establishes the equivalence relation between them.
In Sect. 3 we extend the analysis to include the angular variables into our consideration.
The symmetry underlying this case is $so(1,2)\oplus su(2)$ and we
expose an appropriate extension of the model of Ref.~\cite{aff} which supports this symmetry
and is canonically equivalent to the particle on $AdS_2\times S^2$. Sect. 4 is devoted
to possible applications of the canonical transformation we found. In particular,
we construct a Hamiltonian of a $N=4$ superparticle on $AdS_2\times S^2$
by firstly supersymmetrizing our simple conformal model and then applying
the canonical transformation to the resulting system. Some open questions
and further developments are discussed in the concluding Sect. 5.

\vspace{0.4cm}

\noindent
{\bf 2. $AdS_2$ background as a canonical transformation of conformal mechanics}\\

\noindent The motion of a charged massive particle near the horizon of an extreme
Reissner-Nordstr\"om black hole is governed by the (static gauge) action functional
\be\label{rn}
S=\int d t {(2R/r)}^2 \left[ q-\m\sqrt{1-{(r/2R)}^2 {\dot r}^2 -R^2 {(r/2R)}^4 ({\dot\t}^2
+\sin^2 \t {\dot\vf}^2)   }~\right].
\ee
Here $\m$ and $q$ stand for the mass and electric charge of the particle and $R$
is the radius of the sphere in the underlying ${AdS}_2 \times S^2$ geometry
(which is equal to that of the $AdS_2$ space and coincides with
the black hole `mass' in units for which $G=1$). As has been argued in Ref.~\cite{kallosh},
in the limit $R\rightarrow \infty$, $(\m-q)\rightarrow 0$, with $R^2(\m-q)$ fixed,
the corresponding quantum mechanical description reduces to that of the `old' (or
`non-relativistic') conformal mechanics~\cite{aff}
\be\label{act}
S=\frac 12 \int d t \left( {\dot x}^2 - \frac{\hat g}{x^2} \right),\quad x=\sqrt{\mu}\, r \;,
\ee
provided $\hat g=8R^2 \mu(\m-q)+4l(l+1)$. Here $l$ stands for the orbital angular momentum
of the particle. This relation between the two models has been recognized to be
a manifestation of the $AdS_2/CFT_1$ correspondence.

Since in the aforementioned limit the angular variables effectively decouple and show up in
an indirect way only in the coupling constant $\hat g$, it seems interesting to discuss a
connection between the radial part of the model~(\ref{rn}) and conformal
mechanics~(\ref{act}) in more detail. According to a recent analysis~\cite{ikn},
for a finite nonzero value of the radius $R$ and $l=0$ the systems are equivalent and correspond
to two different nonlinear realizations of the conformal group $SO(1,2)$.
In particular, the actions~(\ref{rn}) and (\ref{act}) at $ \theta=\varphi=const$ and
$\hat g=8R^2 \mu(\m-q)\equiv g$  are connected by a specific field
redefinition involving coordinates along with their time derivatives.

It turns out that a similar conclusion can be reached in a simpler and suggestive way
if one switches to the Hamiltonian framework. The former case is characterized by
the Hamiltonian
\be\label{adsham}
H_{AdS}={(2R/r)}^2 [\sqrt{\m^2+{(r/2R)}^2 p_r^2 +{(1/R)}^2(p_\t^2+\sin^{-2}\t p_\vf^2)} -q],
\ee
and for our subsequent discussion in this section we will need only the radial part
\be\label{ham2}
H={(2R/r)}^2 \left[ \sqrt{\m^2 +{(r/2R)}^2 p_r^2} -q\right].
\ee
Apart from time translations generated by this Hamiltonian one reveals
two more conserved charges corresponding to dilatations and special conformal transformations
\be
D=tH-\frac 12 r p_r, \quad K=t^2 H-t(r p_r) +\frac 14 r^2 \left(\sqrt{\m^2 +{(r/2R)}^2 p_r^2} +q\right).
\ee
Altogether these form a $so(1,2)$ algebra
\be\label{confalg}
\{H,D \}=H, \quad \{H,K \}=2D, \quad \{D,K \} =K~,
\ee
under the standard Poisson bracket $\{ r, p_r\} = 1$,
which is the conformal algebra in $d=1$. In the conformal mechanics case
\p{act} (with $\hat g=g$) a representation of the algebra reads~\cite{aff}
\be\label{ham1}
H=\frac{1}{2}\left( p^2 + \frac{g}{x^2} \right), \quad
D=tH-\frac 12 xp, \quad K=t^2 H-t(xp) +\frac 12 x^2.
\ee
Searching for a classical correspondence between the two models, we wonder if there exists a transformation from the
phase space coordinates
$(x,p)$ to $(r,p_r)$ which brings the Hamiltonian in Eq.~(\ref{ham1})
to the form~(\ref{ham2}).
Furthermore, since the Hamiltonian makes part of the conformal algebra it seems reasonable
to strengthen the condition and demand {\it all} the conformal generators to coincide.
Comparing the charges corresponding to dilatations one immediately finds
\be
xp=rp_r,
\ee
while requiring the identity of the charges generating special conformal
transformations leads one to set
\be\label{cantr}
x=\frac {1}{\sqrt 2} r{ \left[\sqrt{\m^2 +{(r/2R)}^2 p_r^2} +q\right]}^{\frac 12}, \quad
p=\sqrt 2 p_r {\left[ \sqrt{\m^2 +{(r/2R)}^2 p_r^2} +q \right]}^{-\frac 12}.
\ee
It is straightforward to verify that, being performed in the Hamiltonian~(\ref{ham1}),
this substitution does produce Eq.~(\ref{ham2}), provided the identification
$g={(2R)}^2 (\m^2 -q^2)$. Notice that this correlates well with the coupling constant
appearing in the aforementioned limit
\be
g=(2R)^2( \mu^2-q^2) \rightarrow 8R^2\mu (\mu - q)~,
\ee
if one suppresses the angular variables. Besides, the transformation~(\ref{cantr})
is {\it canonical} with the unit Jacobian.

We thus demonstrated that at the classical level the radial part of a charged massive
particle moving near the horizon of an extreme Reissner-Nordstr\"om black hole is canonically
equivalent to the old conformal mechanics. This equivalence is implicit in
the Hamiltonian analysis of Ref.\cite{zanon}. In the above, we established this connection
in an explicit way. Moreover, the method by which we have reached this conclusion, i.e.
the principle of identifying the symmetry generators, is new and allows one to treat
more complicated cases (see next sections).

It is worth mentioning that according to the analysis of Ref.~\cite{zanon}
(see also references therein) the system~(\ref{act}) in the Hamiltonian approach exhibits
a larger symmetry than one could expect to find. In particular, it was shown that
the $so(1,2)$ algebra formed by the conserved charges $H,D,K$ can be extended to
$w_\infty$ algebra of area-preserving symplectic diffeomorphisms, the latter including
the Virasoro algebra as a subalgebra. It was subsequently realized~\cite{cad}, however,
that the charges are functionally dependent which matches with the fact
that the system~(\ref{act}) involves only a finite number of degrees of freedom.
Due to the existence of the equivalence transformation (\ref{cantr}) the same symmetries should also
persist in the model with the Hamiltonian (\ref{adsham}).
In what follows we shall concentrate only on finite dimensional subalgebras.

\vspace{0.5cm}

\noindent
{\bf 3. Adding angular variables}

\vspace{0.5cm}

\noindent Guided by the observation made in the preceding section it seems natural to inquire whether
it is possible to extend the conformal mechanics~(\ref{act}) by angular variables so as to
construct a model canonically equivalent to the particle moving on the ${AdS}_2 \times S^2$
background. A reasonably good starting point is offered by the Hamiltonian
\be\label{ham4}
H=\frac{1}{2}\left( p^2 + \frac{g}{x^2} \right)+\frac{2}{x^2} (p_\T^2+\sin^{-2} \T p_\Phi^2),
\ee
which exhibits conformal symmetry (the generators of dilatations and special conformal
transformations maintain their form~(\ref{ham1}) with $H$ defined by Eq.~(\ref{ham4}))
along with the rotation $SO(3)$ invariance. The Hamiltonian (\ref{ham4})
arises from (\ref{adsham}) in the same limit $R\rightarrow \infty$,
$(\m-q)\rightarrow 0$, and $R^2(\m-q)$ fixed,  with the full angular part
being taken into account. Now we are going to demonstrate that it produces
(\ref{adsham}) after performing a proper canonical transformation (with
$g=(2R)^2( \mu^2-q^2)$) .

For the model at hand a representation of the $su(2)$ algebra is realized in the
standard way ($\epsilon_{123}=1$)
\bea\label{su2}
&&
{{\mathcal{J}}_1}=-p_\Phi \cot \T \cos \Phi  -p_\T \sin \Phi, \quad
{{\mathcal{J}}_2}=-p_\Phi \cot \T \sin \Phi  +p_\T \cos \Phi ,
\nonumber\\[2pt]
&& \qquad \qquad \qquad \qquad  \qquad
{{\mathcal{J}}_3}=p_\Phi; \quad  \{ {{\mathcal{J}}_i},{\mathcal{J}}_j \}=
\epsilon_{ijk} {{\mathcal{J}}_k},
\eea
and it is noteworthy that the angular part of the Hamiltonian is provided by the
Casimir operator of the $su(2)$ algebra ${\mathcal{J}}^2={\mathcal{J}}_i {\mathcal{J}}_i
={p_\T}^2 +\sin^{-2} \T {p_\Phi}^2$.

Much alike the preceding case a transformation
$(x,\T,\Phi,p,p_\T,p_\Phi) \rightarrow (r,\t,\vf,p_r,p_\t,p_\vf)$
which brings the test Hamiltonian~(\ref{ham4}) to that associated with the model~(\ref{rn})
(see Eq.~(\ref{adsham}) above) is relatively easy to deduce for the radial variables
by comparing the relevant expressions for the conformal generators
\bea\label{tran}
&&
x=\frac{1}{\sqrt{2}} r{[\sqrt{\m^2+{(r/2R)}^2 p_r^2 +{(1/R)}^2(p_\t^2+\sin^{-2} \t p_\vf^2)} +q]}^{1/2},
\nonumber\\[2pt]
&&
p=\sqrt{2}p_r{[\sqrt{\m^2+{(r/2R)}^2 p_r^2 +{(1/R)}^2(p_\t^2+\sin^{-2}\t p_\vf^2)} +q]}^{-1/2}.
\eea
Besides, one has to make the identification $q^2=\m^2-g/{(2R)}^2$ and require
the Casimir operator to remain invariant
${p_\T}^2 +\sin^{-2} \T {p_\Phi}^2={p_\t}^2 +\sin^{-2} \t {p_\vf}^2$. The latter requirement,
however, does not fix the canonical transformations for the rest of the involved variables.
Clearly, the reason lies in the additional $SO(3)$ symmetry characterizing the
case under consideration. A sensible way out is to require {\it all} the symmetry generators
in both pictures to coincide. In particular, we put ${\mathcal{J}}_i=J_i$, where the transformed
generators $J_i$ have the same form as in Eq.~(\ref{su2}) but involve $(\t,\vf,p_\t,p_\vf)$
instead of $(\T,\Phi,p_\T,p_\Phi)$. Being algebraic equations, these allow one to express
three variables
\bea\label{cot}
&&
p_\T=J_2 \cos{\Phi} -J_1 \sin{\Phi}, \quad \cot \T=-\frac{1}{J_3} \left( J_1 \cos{\Phi} +J_2 \sin{\Phi} \right),
\quad p_\Phi=p_\vf,
\eea
in terms of $\Phi$. Besides, we demand the change to be canonical.
Let us discuss the latter point in more detail.

The dependence of $\Phi$ on the radial coordinates $(r,p_r)$ is dictated by the requirement
that it commutes with the pair $(x,p)$ from Eq.~(\ref{tran}). Given the transformation~(\ref{tran}),
the equality $xp=rp_r$ holds and one immediately faces the restriction
\be\label{supl}
\{\Phi,rp_r \}=0.
\ee
It means, in particular, that $\Phi$ is a function of $(rp_r)$.
Making use of Eq.~(\ref{supl}) one can verify that only one of the
two equations $\{\Phi,x \}=\{\Phi,p \}=0$ is independent and amounts to
\bea\label{eq2}
&&
\frac{\partial \Phi}{\partial p_r}+
\frac{r\{{\mathbf{J}^2}, \Phi \}}
{{(2R)}^2 \sqrt{\m^2+{(r/2R)}^2 p_r^2 +{(1/R)}^2 {\mathbf{J}^2}}
\left(\sqrt{\m^2+{(r/2R)}^2 p_r^2 +{(1/R)}^2 {\mathbf{J}^2}}+q\right)}
=0,
\nonumber\\[2pt]
&&
\eea
where ${\mathbf{J}}^2=J_i J_i$. Taking into account that the Casimir operator
maintains its form, the explicit expression for  $\{{\mathbf{J}^2}, \Phi \}$
can be easily computed. Then, introducing a specific subsidiary function
\bea\label{sub}
&&
\a=A+\frac{\sqrt{{\mathbf{J}^2}}}{R\sqrt{\m^2 +{(1/R)}^2 {\mathbf{J}^2} -q^2}}
\left[\arctan \left( \frac{rp_r}{2R\sqrt{\m^2 +{(1/R)}^2 {\mathbf{J}^2} -q^2}} \right) \right.-
\nonumber\\[2pt]
&& \qquad
-\arctan \left( \frac{q rp_r}{2R\sqrt{\m^2 +{(1/R)}^2 {\mathbf{J}^2} -q^2}} \right. \left.\left.
\frac{1}{\sqrt{\m^2 +{(r/2R)}^2 p_r^2+{(1/R)}^2 {\mathbf{J}^2}}} \right) \right],
\eea
where $A$ depends on the angular variables $(\t,\vf,p_\t,p_\vf)$
only, one can readily integrate the radial equation~(\ref{eq2})
\bea\label{tran1}
\tan\Phi=\frac{J_3\sqrt{{\mathbf{J}^2}}}{J_2^2+J_3^2}\tan\a-\frac{J_1 J_2}{J_2^2+J_3^2}~.
\eea
Here we made use of Eq.~(\ref{cot}) and assumed the conditions
$\{\Phi, \T\}=\{\Phi, p_\T \}=0, \{\Phi,p_\Phi \}=1$ to hold.
Obviously, the last three equations are designed to fix the explicit form of $A$ which enters the subsidiary function.
A straightforward calculation reveals the following restrictions
\bea
\{A,J_1\}=0, \quad \{A,J_3 \}=\frac{J_3\sqrt{{\mathbf{J}}^2}}{J_2^2+J_3^2},  \quad
\{A,J_2 \}=\frac{J_2\sqrt{{\mathbf{J}}^2}}{J_2^2+J_3^2}~.
\eea
Beautifully enough, the following solution to the first equation:
\be
A=\arctan\left(\frac{p_\vf \sin^{-2} \t \tan\vf -p_\t \cot\t}{\sqrt{{\mathbf{J}}^2}} \right),
\ee
solves the others as well.

Having specified the explicit form of $\Phi$, one has to verify yet that the whole change is canonical.
It proves to be the case. In particular, the conjugate momentum $p_\Phi$ commutes
with $(x,p,\T,p_\T)$ while the pair $(\T,p_\T)$ is canonical $\{\T,p_\T \}=1$. Besides, as $\Phi$ has the vanishing bracket
with the pair $(x,p)$, so do $(\T,p_\T)$.

To summarize, the canonical change of the variables exposed above in
Eqs.~(\ref{tran}), (\ref{cot}), (\ref{tran1}) establishes the equivalence relation between the
charged massive particle moving near the horizon of an extreme Reissner-Nordstr\"om black hole
(see Eq.~(\ref{adsham}) above) and conformal mechanics~(\ref{ham4}). Although the transformation
looks pretty bulky when applied to the angular variables, the quantities of physical interest
like the angular momentum or the angular contribution to the Hamiltonian remain invariant and
are easily handled. It is noteworthy that the equivalence holds for any fixed value of the black
hole mass and is not bound to any specific limit.

\vspace{0.5cm}

\noindent
{\bf 4. A $N=4$ superparticle on $AdS_2\times S^2$ background}

\vspace{0.5cm}

\noindent Among possible applications of the model~(\ref{ham4}) which we briefly outline in the concluding
section there is one which can be addressed immediately. It has been known for a long
time that conformal mechanics~(\ref{act}) admits supersymmetric
generalizations~\cite{pashnev,fub,ikl}. It is interesting to find
analogous superextensions of the particle on $AdS_2\times S^2$.
Since the full superisometry of the $AdS_2\times S^2$ background is known to be
$SU(1,1|2)$, the corresponding $N=4$ superconformal mechanics
should possess this symmetry.
In this context the $SU(2)$ symmetry underlying the bosonic case
comes out as the $R$--symmetry contained in the superconformal group.

In order to construct a $N=4$ superconformal mechanics in AdS space one could either
use the non-linear realizations~\cite{ikl,azc}, or properly fix the gauge with respect to
$\kappa$-symmetry in the $0$--brane Green-Schwarz action on $AdS_2\times S^2$~\cite{zhou} or,
working in a more general geometric setting, analyse the conditions for a particle moving
in an arbitrary curved background to admit a $N=4$ superconformal symmetry (see
e.g. Refs~\cite{strom,pap,don}). Observe now that our consideration in
the preceding section suggests quite new and interesting possibility to construct
a $su(1,1|2)$--invariant superconformal mechanics in $AdS_2\times S^2$ space by making use of
the Hamiltonian approach. Indeed, it suffices to extend the simple model~(\ref{ham4})
by fermions in a manner which complements the $so(1,2)\oplus su(2)$--symmetry algebra of
the bosonic case
to the entire $su(1,1|2)$ and then apply to the resulting theory the canonical transformation
found above with the fermions kept untouched.

The construction turns out to be mostly algebraic. One introduces a pair of complex fermions
${(\psi^i)}^{*}=\bar\psi_i$, $i=1,2$, obeying the bracket $\{\psi^i,\bar\psi_j \}=-i{\delta^i}_j$,
and modifies the $su(2)$ generators~(\ref{su2}) by adding the appropriate fermionic bilinears
(without spoiling the algebra!)
\bea
&&
\tilde{\mathcal{J}}_1={\mathcal{J}}_1+\frac{i}{2}(\psi^2\bar\psi_1 -\psi^1\bar\psi_2), \quad
\tilde{\mathcal{J}}_2={\mathcal{J}}_2-\frac{1}{2}(\psi^2\bar\psi_1 +\psi^1\bar\psi_2), \quad
\nonumber\\[2pt]
&& \qquad \qquad \qquad
\tilde{\mathcal{J}}_3={\mathcal{J}}_3+\frac{1}{2}(\psi^1\bar\psi_1 -\psi^2\bar\psi_2).
\eea
Requiring them to obey proper Poisson brackets with the Poincar\'e supersymmetry
generators $G^i,\bar G_i$, one severely restricts the form of the latter.
Observing further that the bracket
$\{G^i,\bar G_j \}=-2iH{\delta^i}_j$, $i=1,2$, makes part of the $su(1,1|2)$ superalgebra,
it suffices to find fermionic generators  $G^i$ and $\bar G_i$ whose Poisson
bracket yields a Hamiltonian which reduces to Eq.~(\ref{ham4}) in the bosonic limit.
Besides, one has to make sure that the conditions $\{G^i,G^j\}=\{\bar G_i,\bar G_j\}=0$
hold which, by Jacobi identities, provide the conservation of the supercharges.
It should be also mentioned that, in order to guarantee the stability
of the vacuum (see the discussion in Refs.~\cite{kallosh,azc}) one is forced to set $\mu=q$.
We thus put $g=0$ in our subsequent consideration.

It turns out that all these restrictions are met by the following representation for the
supersymmetry charges
\bea
&&
G^1=\left(p-\frac{2i}{x} {\mathcal{J}}_3 \right)\psi^1+\frac{2}{x}
\left({\mathcal{J}}_1+i{\mathcal{J}}_2\right)\psi^2
+\frac{i}{x}\psi^1 \psi^2 \bar\psi_2,
\nonumber\\[2pt]
&&
G^2=-\left(p+\frac{2i}{x} {\mathcal{J}}_3 \right)\psi^2+\frac{2}{x}
\left({\mathcal{J}}_1-i{\mathcal{J}}_2\right)\psi^1
-\frac{i}{x}\psi^1 \bar\psi_1 \bar\psi^2,
\eea
which yield the Hamiltonian
\bea\label{ham7}
&&
H=\frac 12 \left[p^2 +\frac{4}{x^2} (p_\T^2+\sin^{-2} \T p_\Phi^2) \right]
+\frac{2i}{x^2} ({\mathcal{J}}_1-i{\mathcal{J}}_2)\psi^1\bar\psi_2 -\frac{2i}{x^2}
({\mathcal{J}}_1+i{\mathcal{J}}_2)\psi^2\bar\psi_1
\nonumber\\[2pt]
&& \qquad
-\,\frac{2}{x^2}{\mathcal{J}}_3(\psi^1\bar\psi_1-\psi^2\bar\psi_2)
+\frac{1}{x^2}\psi^1\bar\psi_1\psi^2\bar\psi_2.
\eea
Given the Hamiltonian, one can readily verify that the generators of dilatations and special
conformal transformations maintain their previous form~(\ref{ham1}) (with the Hamiltonian
taken from the previous line). Finally, evaluating the Poisson brackets of the supersymmetry
charges with the generators of special conformal transformations one finds a representation
for the superconformal generators
\be
S^1=tG^1-x\psi^1,\quad  S^2=t G^2+x \psi^2.
\ee

Having formulated the model in the conformal basis, we now proceed to construct its
$AdS_2\times S^2$ equivalent.
To this end we apply the transformation~(\ref{tran}) (with $\mu=q$)
to the Hamiltonian~(\ref{ham7}) which yields
\bea\label{last}
&&
H_{N=4}={(2R/r)}^2 [\sqrt{\m^2+{(r/2R)}^2 p_r^2 +{(1/R)}^2 {\mathbf{J}}^2} -\mu]
+\left[(J_1-iJ_2)\psi^1\bar\psi_2-(J_1+iJ_2)\psi^2\bar\psi_1
\right.
\nonumber\\[2pt]
&&
\left.+
iJ_3(\psi^1\bar\psi_1-\psi^2\bar\psi_2)-\frac{i}{2}
\psi^1\bar\psi_1\psi^2\bar\psi_2\right]
\frac{4i}
{r^2(\sqrt{\m^2+{(r/2R)}^2 p_r^2 +{(1/R)}^2 {\mathbf{J}}^2} +\mu)}.
\eea
Because the bosonic limit of this theory does coincide with the Hamiltonian~(\ref{adsham})
one ends up with a $SU(1,1|2)$ supersymmetric generalization of the model~(\ref{rn}).
It is interesting to compare the result with the $SU(1,1|2)$ superparticle in the Green-Schwarz approach
\cite{zhou}. This requires the construction of a Lagrangian formulation which will be given elsewhere.

\vspace{0.5cm}

\noindent
{\bf 5. Conclusion}

\vspace{0.5cm}

\noindent To summarize, in the present paper we took advantage of the Hamiltonian formalism, in
order to establish a precise classical correspondence between a massive charged particle
moving near the horizon of an extreme Reissner-Nordstr\"om black hole
and conformal mechanics~(\ref{ham4}). Since our construction does not rely upon a
specific limit, it becomes possible to investigate in full generality quantum properties of the
former model (at any finite value of the black hole mass) working with the latter theory.
It is then tempting to study the quantum spectrum and the transition amplitude for the
theory~(\ref{ham4}). Although we have a little hope to literally transform
into the $AdS$ basis the results of the operator quantization because of the complexity of the
transformations~(\ref{tran}), (\ref{cot}), (\ref{tran1}), the path integral
quantization is still quite feasible.

In constructing a $N=4$ supersymmetric generalization of the model~(\ref{ham4})
we assumed the stability of the vacuum and set $g=0$. The case $g\ne 0$ can also be considered.
It also remains to explore how the equivalence in
the Hamiltonian approach is translated into the Lagrangian language and how
it is linked to the off-shell map of Refs. \cite{ikn,bik}.

\vspace{0.5cm}

\noindent{\bf Acknowledgements}\\

\noindent This work was partially supported by INTAS grant No 00-0254 and the Iniziativa
Specifica MI12 of the INFN Commissione Nazionale IV, as well as by
European Community's Human Potential Programme contract HPRN-CT-2000-00131 (S.B.),
NATO Collaborative Linkage Grant PST.CLG.979389 (S.B. and A.G.), grants DFG No.436
RUS 113/669, RFBR-DFG 02-02-04002, RFBR-CNRS 01-02-22005 and a grant of the Heisenberg-Landau
program (E.I. and S.K.).

\end{document}